\documentclass[12pt]{article}
\usepackage{amsfonts,latexsym,graphicx,amsmath}
\usepackage[colorlinks]{hyperref}
\hypersetup{citecolor=red}
\hypersetup{linkcolor=blue}
\hypersetup{urlcolor=blue}
\usepackage{tikz}
\usepackage{doi}
\usepackage{xcolor}
\usepackage[font=small]{caption}

\title{Bethe Ansatz solution for a model of global-range \\ interacting bosons on the square lattice}

\author{Jon Links \\
School of Mathematics and Physics, \\
The University of Queensland, 4072, \\
Australia}

\begin{document}
\maketitle

\abstract{Quantum systems on a one-dimensional lattice are ubiquitous in the study of  models exactly-solved by Bethe Ansatz techniques. Here it is shown that including global-range interaction opens scope for Bethe Ansatz solutions that are not constrained to one-dimensional quantum systems. A bosonic model on a square lattice is defined, and the exact Bethe Ansatz solution is provided for open, cylindrical, and toroidal boundary conditions. Generalising the result for an integrable defect leads to a Bethe Ansatz solution that is not expressible in an exact, closed-form manner.  
}

\section{Introduction}
\label{sec:1}

For each of the fundamental particle properties known as spin, boson, and fermion, there exists a fundamental, one-dimensional model accommodating a Bethe Ansatz solution. These are, respectively, the Heisenberg spin-1/2 model \cite{b31}, the Bose gas \cite{ll63}, and the Hubbard model \cite{lw68}. While the Heisenberg and Hubbard models are defined on a lattice, the Bose gas Hamiltonian is formulated through continuous field operators. The natural discretised version of the Bose gas, the Bose-Hubbard model, is generally considered to not be integrable \cite{bmmbk20,fp03,krbl10,k16,kb04,nh23,ol07}. While there have been attempts to formulate alternative lattice discretisations of the Bose gas model, these have generally possessed complicating features, such as requiring intricate alternating-site expressions \cite{bk86,ik84}, or being non-Hermitian \cite{eb24,kr94,zqwcc24}. Moreover, while there are well-developed methods for generating and/or testing  integrability in models for spin \cite{dpprr21,gm95,h25,h25b,s19,z25} or fermionic \cite{dprr20,lzmg01} degrees of freedom using boost (a.k.a. ladder) operator techniques and the Reshetikhin criterion, these do not readily extend to bosonic systems. Consequently, examples of exactly-solved bosonic lattice models are infrequently encountered.     

Here, a Hamiltonian is introduced describing bosons on a square lattice with nearest-neighbour hopping, and a global-range (a.k.a. infinite-range) collective interaction term. The form of this interaction was proposed in \cite{lhdlmde16} to model the interaction mediated by an optical cavity in an experiment using $^{87}$Rb cold atoms confined to a stack of square optical lattices. 
For a $2\times 2$ four-site lattice, the same form of interaction appears in \cite{gwylf22} for a system of $^{164}$Dy atoms with  
long-ranged magnetic dipole interactions, discussed in relation to NOON state formation; see also \cite{bil23,gywtfl22}. Other studies involving global-range interactions include \cite{cblmz25,dbhd16,hdlde18} for bosonic models, and \cite{ibrr18,osirf20,zz24} for spin models.

There has been a resurgence of interest in exactly-solvable models with long-range interactions in recent times; e.g. \cite{blst25,dr23,g22,kl24}. These works consider the systems to be one-dimensional spin chains. However, once long-range interactions are incorporated there is a decreased clarity in the notion of dimensionality. A case in point is provided by the Haldane-Shastry model \cite{h88,s88}, the integrability of which does not stem from the typical transfer matrix approach associated with the Yang-Baxter equation \cite{bghp93}. The model is generally considered to be one-dimensional, but the geometry of the model can be viewed as being a ring with periodic boundary conditions embedded in two-dimensional Euclidean space. This is due to the strength of spin interactions in the Hamiltonian being dependent on the {\it chord distance} between sites on the ring \cite{h88}.      

In the next section the model with global-range interaction is formulated in terms of canonical boson operators associated with the sites of a square lattice. A key element in the analysis is the introduction of the adjacency matrix to describe the Hamiltonian. 
The Bethe Ansatz solution for the Hamiltonian of this study is obtained by undertaking a canonical transformation to a spectrally equivalent system with hopping on a collection of disjointed lattices. This simplifies the form of the Hamiltonian, and allows the solution to be read off from previously published results.    
The case of open, cylindrical, and toroidal boundary conditions are studied. However it will be seen that this approach, developed for bosons on the square lattice, bears little resemblance to the typical transfer matrix method used for spin systems; e.g. \cite{ddd22}.
Finally, the case of an integrable boundary defect is discussed.

\section{Integrable model with global-range interaction}
\label{sec:2}

The hopping interactions of the Hamiltonian will be formulated on an $L\times L$ square lattice, where $L$ is assumed to be even. The approach taken is aided by implementing the notion of an {\it adjacency matrix}, and by exploiting the {\it bipartite} nature of the lattice \cite{illw25}. 
Let $\{a_j,\,a_j^\dagger:j=1,\dots,m\} \cup 
\{b_j,\, b_j^\dagger:j=1,\dots,m\}$, where $m=L^2/2$, denote mutually commuting sets of canonical boson operators satisfying
\begin{align*}
&[a_j,\,a_k^\dagger]=[b_j,\,b_k^\dagger]=\delta_{jk}I, \\
&[a_j,\,a_k]=[a_j^\dagger,\,a_k^\dagger]=[b_j,\,b_k]=[b^\dagger_j,\,b_k^\dagger]=0.
\end{align*}
For adjacency matrix of the form 
\begin{align}
A=
\left(
\begin{array}{ccc}
0 & | & B \\
- & & - \\
B & | & 0 
\end{array}
\right) 
\label{am}
\end{align}
where the submatrix $B$ is a real, symmetric matrix, the Hamiltonian reads
\begin{align}
H&=U(\hat{N}_a-\hat{N}_b)^2+\sum_{j,k=1}^m B_{jk}(a_j^\dagger b_k + b_j^\dagger a_k)
\label{ham}
\end{align}
where
$ \displaystyle
\hat{N}_a=\sum_{j=1}^m a_j^\dagger a_j,\, \hat{N}_b=\sum_{j=1}^m b_j^\dagger b_j$.

The first term on the right-hand side of (\ref{ham}) is precisely the form of interaction used 
to model the effect of an optical cavity applied to bosonic cold atoms in an optical square lattice \cite{cblmz25,dbhd16,hdlde18,lhdlmde16}. The second term of (\ref{ham}) describes the hopping of bosons between neighbouring sites. The two sets of creation and annihilation operators are used to highlight the bipartite property of a square lattice. The vertices of the lattice are partitioned into two non-intersecting sets, 
${\mathcal A}$ and ${\mathcal B}$. For a vertex belonging to ${\mathcal A}$ all nearest-neighbour vertices belong to ${\mathcal B}$, and vice versa. The entry $B_{jk}=B_{kj}$ is equal to one if the $j$th vertex of set ${\mathcal A}$ is a nearest-neighbour of the $k$th vertex of set ${\mathcal B}$; otherwise, $B_{jk}=B_{kj}$ is set to be zero.
This formulation can be applied to any bipartite lattice, e.g. hexagonal, cubic. The focus here will be on square lattices, with different forms of boundaries. 

Note that $\hat{N}=\hat{N}_a+\hat{N}_b$ commutes with (\ref{ham}), indicating that the Hamiltonian conserves total particle number. In addition, there exists a set of conserved operators that follow from the 
Yang-Baxter integrability of (\ref{ham}), which have been derived in \cite{illw25}.
The conserved operators satisfying $[C(y),\,C(z)]=0$, with $y,\,z\in {\mathbb Z}_{\geq 0}$, are given by 
\begin{align}
C(2p)&= \sum_{j,k=1}^m{ B}^{2p}_{jk}(a_j^\dagger a_k + b_j^\dagger b_k ), 
\label{con1}
\\
C(2p+1)&= 2U\sum_{i=0}^{2p} D(2p,i) +\sum_{j,k=1}^m{B}^{2p+1}_{jk}(a_j^\dagger b_k + b_j^\dagger a_k )
\label{con2}
\end{align}
with
\begin{align*}
D(2p,i)=\begin{cases} \displaystyle
\sum_{j,k,r,q=1}^m {B}^i_{jk} {B}^{2p-i}_{rq}(a^\dagger_j a_q a^\dagger_r a_k +
b^\dagger_j b_q b^\dagger_r b_k), \qquad i\,\,{\rm even},  \\
\displaystyle
\sum_{j,k,r,q=1}^m ({B}^i_{jk} {B}^{2p-i}_{rq}+{B}^{2p-i}_{jk} {B}^{i}_{rq}) 
a^\dagger_j a_q b^\dagger_r b_k, \qquad i\,\,{\rm odd}.
\end{cases}
\end{align*}
Above, $B^q_{jk}$ denotes the $(j,k)$ entry of the matrix power $B^q$.
We have $\hat{N}=\hat{N}_a+\hat{N}_b=C(0)$ and  $H=C(1)-U(\hat{N}^2+2(m-1)\hat{N})$, as seen through
\begin{align*}
&C(1)-U(\hat{N}^2+2(m-1)\hat{N}) \\
&\qquad =2U\sum_{j,q=1}^m (a^\dagger_j a_q a^\dagger_q a_j +
b^\dagger_j b_q b^\dagger_q b_j)    +\sum_{j,k=1}^m{B}_{jk}(a_j^\dagger b_k + b_j^\dagger a_k )  -U((\hat{N}_a+\hat{N}_b)^2+2(m-1)\hat{N}) \\
&\qquad =2U(\hat{N}_a(mI+\hat{N}_a)+\hat{N}_b(mI+\hat{N}_b))- 2U\sum_{j,q=1}^m \delta_{jq} (a^\dagger_j a_q  +
b^\dagger_j b_q)  \\
&\qquad \qquad   +\sum_{j,k=1}^m{B}_{jk}(a_j^\dagger b_k + b_j^\dagger a_k )  -U((\hat{N}_a+\hat{N}_b)^2+2(m-1)\hat{N}) \\
&\qquad =U(2\hat{N}_a^2 +2 \hat{N}_b^2 +2 m\hat{N})- 2U\hat{N}  +\sum_{j,k=1}^m{B}_{jk}(a_j^\dagger b_k + b_j^\dagger a_k )  -U((\hat{N}_a+\hat{N}_b)^2+2(m-1)\hat{N}) \\
&\qquad =U(\hat{N}_a - \hat{N}_b)^2 +\sum_{j,k=1}^m{B}_{jk}(a_j^\dagger b_k + b_j^\dagger a_k ) . 
\end{align*}

As mentioned, the study below will only consider cases where the adjacency submatrix $B$ has entries equal to zero or one. However, this is not the most general setting. The operators $C(y)$ are mutually commuting for {\it any} real, symmetric matrix $B$, as a result of Yang-Baxter integrability  \cite{illw25}.

\subsection{Bethe Ansatz solution}
\label{sec:bae}

Obtaining the Bethe Ansatz solution is facilitated by mapping the Hamiltonian (\ref{ham}) to an equivalent model, via a unitary  transformation. Let $X$ denote a unitary matrix that diagonalises $B$, viz.
\begin{align*}
\sum_{p=1}^{m} X^\dagger_{jp} X_{pk} = \delta_{jk}, \qquad
\sum_{p,q=1}^{m} X^\dagger_{jp}B_{pq} X_{qk} = \varepsilon_j \delta_{jk}, 
\end{align*}
with $\{\varepsilon_j:j=1,...,m\}$ denoting the spectrum of 
$B$.
Introducing another pair of mutually commuting sets of canonical boson operators
 $\{c_j,\,c_j^\dagger:j=1,\dots,m\} \cup 
\{d_j,\, d_j^\dagger:j=1,\dots,m\}$,   
satisfying
\begin{align*}
a_{k}=\sum_{j=1}^m X_{kj} c_j, \quad
b_{k}=\sum_{j=1}^m X_{kj} d_j, \quad
a^\dagger_{k}=\sum_{j=1}^m X^\dagger_{jk} c^\dagger_j, \quad
b^\dagger_{k}=\sum_{j=1}^m X^\dagger_{jk} d^\dagger_j, 
\end{align*}
leads to  $\displaystyle \hat{N}_a=\sum_{j=1}^m c_j^\dagger c_j=\hat{N}_c$,   
$\displaystyle \hat{N}_b=\sum_{j=1}^m d_j^\dagger d_j=\hat{N}_d$, and  
\begin{align} 
H&=U(\hat{N}_c-\hat{N}_d)^2 + \sum_{j=1}^m \varepsilon_j (c_j^\dagger d_j + d_j^\dagger c_j).
\label{tham} 
\end{align}
Note that 
$\displaystyle
\hat{N}_j = c^\dagger_j c_j + d^\dagger_j d_j
$
are conserved operators; let $N_j$ denote their eigenvalues. Then
$\displaystyle
\sum_{j=1}^m N_j=N
$
gives the total number of particles.

The effect of the transformation is that takes the square lattice model and maps it to the spectrally equivalent model (\ref{tham}). The transformed model is defined on the union of disconnected graphs, each with only two vertices and one edge. These graphs are weighted by 
$\varepsilon_j$, the coupling for the tunneling strength between the two connected vertices. Thus, the dependence on the matrix $B$ is encoded in these eigenvalues.  However observe that the form of the interaction term with coupling $U$ is invariant under the transformation. 
A graphical representation the transformed Hamiltonian in some examples is provided in panel (c) of Figs. \ref{fig1}, \ref{fig2}, and \ref{fig3} below.

Expressions for the energy eigenvalues of (\ref{tham}) are known from \cite{l17},  using a functional Bethe Ansatz approach for a six-vertex solution of the classical Yang-Baxter equation without skew-symmetry. The energies read  
\begin{align}
E=UN^2 +4U \sum_{j=1}^m  \sum_{n=1}^N  \frac{N_j \varepsilon_j^2}{v_n-\varepsilon_j^2}
\label{nrg}
\end{align}
subject to the parameters $\{v_n:n=1,\dots, N  \}$ satisfying the  Bethe Ansatz equations   
\begin{align}
\sum_{p\neq n}^N \frac{2v_n}{v_n-v_p}
+\frac{\displaystyle \prod_{j=1}^m(v_n-\varepsilon_j^2)^{N_j}}{\displaystyle 16U^2\prod_{p\neq n}^N(v_n-v_p)}
&=N-1+\sum_{j=1}^m \frac{N_j\varepsilon_j^2}{v_n-\varepsilon^2_j}
\label{bae}
\end{align}
for $n=1,\dots,N$. Note that both (\ref{nrg}) and (\ref{bae}) are functions of the {\it squares} of the 
$\{\varepsilon_j:j=1,...,m\}$. This is due to a symmetry resulting from unitary transformations of the form $c_j\mapsto -c_j$ which induces a sign change for the $\varepsilon_j$ in (\ref{tham}). 

The Bethe eigenstates of the transformed operator (\ref{tham}) arise as a generalisation of results presented in \cite{cdv16}. Alternatively, they may be obtained as a specialisation after a transformation (e.g. see \cite{vl25}) of a more recent construction developed in \cite{s23}. A distinctive character of the Bethe eigenstate formulation is that it is described through a set of pseudovacua, rather than a single pseudovacuum, analogous to the constructions in \cite{ytfl17}. 
For a given collection of quantum numbers $\{N_1,\,\dots,\,N_m: N_j\in\,{\mathbb Z}_{\geq 0}\}$, denoting the eigenvalues of the conserved operators 
$\hat{N}_j$ as before, set 
\begin{align*}
|N_1,\dots,\,N_m\rangle= (d_1^\dagger)^{N_1}\dots (d_m^\dagger)^{N_m}|0\rangle
\end{align*}     
where $|0\rangle$ denotes the physical vacuum. For each index $j$, the set $\{c_j^\dagger c_j,\, d_j^\dagger d_j,\,c_j^\dagger d_j,\,d_j^\dagger c_j\}$ closes under the commutator to realise the $gl(2)$ Lie algebra. Each of the pseudovacua $|N_1,\dots,\,N_m\rangle$ is annihilated by the $d_j^\dagger c_j$, since the physical vacuum $|0\rangle$ is annihilated by the $c_j$.
For a solution of the Bethe Ansatz equations (\ref{bae}) the associated Bethe eigenstate reads 
\begin{align*}
|v_1,\dots,v_N;N_1,\dots,\,N_m\rangle = \prod_{n=1}^N {\mathcal C}(v_n) |N_1,\dots,\,N_m\rangle
\end{align*}
where 
\begin{align*}
{\mathcal C}(u)=\frac{1}{2U}I+   \sum_{j=1}^m \frac{2\varepsilon_j}{u-\varepsilon_j^2}  c_j^\dagger d_j .
\end{align*}

\section{Square lattice models}

\subsection{Open boundary conditions}

To illustrate the process for constructing the models, first consider Fig. \ref{fig1} (a) depicting a $4\times 4$ {\it open} lattice. The sites of the lattice are coloured blue or red, in such a way that there are no edges between sites of the same colour. Viewed as a graph, Fig. \ref{fig1} (a) is topologically equivalent to Fig. \ref{fig1} (b) in that they have the same edge sets and vertex (i.e. site) sets. From  Fig. \ref{fig1} (b) the adjacency  submatrix $B$ is read off as follows.  
Assign the labels $1,\dots,8$ to the blue sites, and also to red sites, from top to bottom. Then assign the matrix element $B_{jk}$ to be equal to 1 if there is an edge between the $j$th blue site and the $k$th red site, and equal to zero otherwise. For the instance of Fig. \ref{fig1} (b) this produces the symmetric matrix   
\begin{align}
B=
\left(
\begin{array}{cccccccc}
0 & 1 & 1 & 0 & 0 & 0 & 0 & 0 \\
1 & 1 & 0 & 1 & 0 & 0 & 0 & 0 \\
1 & 0 & 0 & 1 & 1 & 0 & 0 & 0 \\
0 & 1 & 1 & 1 & 0 & 1 & 0 & 0 \\
0 & 0 & 1 & 0 & 0 & 1 & 1 & 0 \\
0 & 0 & 0 & 1 & 1 & 1 & 0 & 1 \\
0 & 0 & 0 & 0 & 1 & 0 & 0 & 1 \\
0 & 0 & 0 & 0 & 0 & 1 & 1 & 1 
\end{array}
\right), 
\label{bmatrixobc}
\end{align}
while the associated $16\times 16$ matrix given through (\ref{am}) 
is the adjacency matrix expressed in a basis that clearly displays the bipartite nature of the lattice.

\begin{figure}[h]
\centering
\captionsetup{width=.85\linewidth}
\begin{tikzpicture}[scale=.53,auto=left,every node/.style={circle,scale=0.6}]

\node[circle, fill=blue] (n31) at (1,6) {};
\node[circle, fill=blue] (n32) at (3,4) {};
\node[circle, fill=blue] (n33) at (1,2) {};
\node[circle, fill=blue] (n34) at (3,0) {};
\node[circle, fill=blue] (n51) at (5,6) {};
\node[circle, fill=blue] (n52) at (7,4) {};
\node[circle, fill=blue] (n53) at (5,2) {};
\node[circle, fill=blue] (n54) at (7,0) {};
\node[circle, fill=red] (n41) at (3,6)  {};
\node[circle, fill=red] (n42) at (1,4)  {};
\node[circle, fill=red] (n43) at (3,2)  {};
\node[circle, fill=red] (n44) at (1,0)  {};
\node[circle, fill=red] (n61) at (7,6)  {};
\node[circle, fill=red] (n62) at (5,4)  {};
\node[circle, fill=red] (n63) at (7,2)  {};
\node[circle, fill=red] (n64) at (5,0)  {};

\foreach \from/\to in {n31/n41,n41/n51,n51/n61,n32/n42,n42/n32,n32/n62,n52/n62,n33/n43,n43/n53,n53/n63,n31/n42,n42/n33,n41/n32,n32/n43,n51/n62,n62/n53,n61/n52,n52/n63,n54/n64,n44/n34,n34/n64,n34/n43,n54/n63,n33/n44,n53/n64}
  \draw[line width=1.5pt] (\from) -- (\to);

\node[circle, fill=blue] (n11) at (9,7) {};
\node[circle, fill=blue] (n12) at (9,6) {};
\node[circle, fill=blue] (n13) at (9,5) {};
\node[circle, fill=blue] (n14) at (9,4) {};
\node[circle, fill=blue] (n15) at (9,3) {};
\node[circle, fill=blue] (n16) at (9,2) {};
\node[circle, fill=blue] (n17) at (9,1) {};
\node[circle, fill=blue] (n18) at (9,0) {};
\node[circle, fill=red] (n21) at (12,7)  {};
\node[circle, fill=red] (n22) at (12,6)  {};
\node[circle, fill=red] (n23) at (12,5)  {};
\node[circle, fill=red] (n24) at (12,4)  {};
\node[circle, fill=red] (n25) at (12,3)  {};
\node[circle, fill=red] (n26) at (12,2)  {};
\node[circle, fill=red] (n27) at (12,1)  {};
\node[circle, fill=red] (n28) at (12,0)  {};

\foreach \from/\to in {n11/n22,n11/n23,n12/n22,n12/n21,n12/n24,n13/n21,n13/n24,n13/n25,n14/n22,n14/n23,n14/n24,n14/n26,n15/n23,n15/n26,n16/n24,n16/n25,n16/n26,n18/n26,n18/n27,n18/n28,n16/n28,n17/n25,n17/n28,n15/n27}
 \draw[line width=1.5pt] (\from) -- (\to);

\node[circle, fill=teal] (p11) at (14,7) {};
\node[circle, fill=teal] (p12) at (14,6) {};
\node[circle, fill=teal] (p13) at (14,5) {};
\node[circle, fill=teal] (p14) at (14,4) {};
\node[circle, fill=teal] (p15) at (14,3) {};
\node[circle, fill=teal] (p16) at (14,2) {};
\node[circle, fill=teal] (p17) at (14,1) {};
\node[circle, fill=teal] (p18) at (14,0) {};
\node[circle, fill=purple] (p21) at (17,7)  {};
\node[circle, fill=purple] (p22) at (17,6)  {};
\node[circle, fill=purple] (p23) at (17,5)  {};
\node[circle, fill=purple] (p24) at (17,4)  {};
\node[circle, fill=purple] (p25) at (17,3)  {};
\node[circle, fill=purple] (p26) at (17,2)  {};
\node[circle, fill=purple] (p27) at (17,1)  {};
\node[circle, fill=purple] (p28) at (17,0)  {};

\foreach \from/\to in {p11/p21,p12/p22,p13/p23,p14/p24,p15/p25,p16/p26,p17/p27,p18/p28}
 \draw[magenta, line width=1.5pt] (\from) -- (\to);

\draw (4,-1.1)     node[scale=1.8]    {(a)};
\draw (10.5,-1.1)  node[scale=1.8]  {(b)};
\draw (15.5,-1.1)  node[scale=1.8]   {(c)};

\end{tikzpicture}

\caption{Schematic representation of the hopping terms in the Hamiltonian (\ref{ham}) for open boundary conditions. (a) Example using the $4\times 4$ square lattice, with sites coloured red and blue to highlight the bipartite structure. The 4 corner vertices have 2 edges, the remaining 8 boundary vertices have 3 edges, while the 4 internal vertices have 4 edges. (b) Equivalent bipartite graph representation that allows the adjacency submatrix $B$ to be determined, as given by Eq. (\ref{bmatrixobc}). Observe that in this representation there are 4 vertices that have 2 edges (1st left, 1st right, 7th left, 7th right), 4 vertices that have 4 edges
(4th left, 4th right, 6th left, 6th right), while the remaining 8 vertices have 3 edges. (c) Transformed bipartite graph representation resulting from the diagonalisation of $B$. This results in a union of disconnected, two-vertex bipartite graphs, with the eigenvalues of $B$ weighting the edges. This leads to the Bethe Ansatz solution (\ref{nrg},\ref{bae}).}

\label{fig1}
\end{figure}
The eigenvalues for (\ref{bmatrixobc}) are found to be given by the multiset 
\begin{align*}
\{-\sqrt{5},1-\sqrt{5},0,0,1,1,\sqrt{5},1+\sqrt{5}\}.
\end{align*}
Since (\ref{nrg}), (\ref{bae}) only depend on the {\it squares} of these eigenvalues, we can equally take the multiset
\begin{align}
{\mathcal S}^4_{{\rm o}}=\{0,0,1,1,\sqrt{5}-1,\sqrt{5},\sqrt{5},1+\sqrt{5}\}.
\label{setobc}
\end{align}

The illustrative example above generalises to the case of an arbitrary $L\times L$ lattice with open boundary conditions. The eigenvalues of the adjacency matrix  $A$ can be directly obtained via Fourier series - see Appendix. The eigenvalues of $A$ have the form
\begin{align}
2\cos\left(\frac{\pi j}{L+1}\right)+2\cos\left(\frac{\pi k}{L+1}\right) \qquad  
j,k\in\{1,\dots,L\} .
\label{formulaobc}
\end{align}
The above formula provides $L^2$ values. 
Note that $A$ as given by (\ref{am}) is spectrally equivalent to 
\begin{align*}
\left(
\begin{array}{ccc}
B & | & 0 \\
- & & - \\
0 & | & -B 
\end{array}
\right). 
\end{align*}
Hence the eigenvalues of $A^2$ occur in (at least) two-fold degenerate pairs. In applying 
(\ref{nrg},\ref{bae}) each degeneracy must be halved to obtain the $m=L^2/2$ solutions for the eigenvalues $\{\varepsilon_j^2\:j=1,...,m\}$ of $B^2$. For example, the positive values and half of the zero values coming from  Eq. (\ref{formulaobc}) for $L=4$ exactly coincides with (\ref{setobc}).

\newpage
\subsection{Cylindrical boundary conditions}

If an analogous model is considered, but with {\it cylindrical} boundary conditions, the only difference in the formulation of the Hamiltonian comes via additional entries in the adjacency matrix. A pictorial representation of the $4\times 4$ lattice is provided in Fig. \ref{fig2} (a). The adjacency submatrix $B$ may be read off from the graph in Fig. \ref{fig2} (b), which is topologically equivalent to Fig. \ref{fig2} (a). It reads  
\begin{align}
B=
\left(
\begin{array}{cccccccc}
0 & 1 & 1 & 0 & 0 & 0 & {\mathbf 1} & 0 \\
1 & 1 & 0 & 1 & 0 & 0 & 0 & {\mathbf 1} \\
1 & 0 & 0 & 1 & 1 & 0 & 0 & 0 \\
0 & 1 & 1 & 1 & 0 & 1 & 0 & 0 \\
0 & 0 & 1 & 0 & 0 & 1 & 1 & 0 \\
0 & 0 & 0 & 1 & 1 & 1 & 0 & 1 \\
{\mathbf 1} & 0 & 0 & 0 & 1 & 0 & 0 & 1 \\
0 & {\mathbf 1} & 0 & 0 & 0 & 1 & 1 & 1 
\end{array}
\right) 
\label{bmatrixcbc}
\end{align}
Note that 
Fig. \ref{fig2} (b) differs from Fig. \ref{fig1} (b) by the inclusion of 4 additional edges. This leads to four additional non-zero entries in $B$, depicted by boldface in Eq. (\ref{bmatrixcbc}).

\begin{figure}[t]
\centering
\captionsetup{width=.85\linewidth}
\begin{tikzpicture}[scale=.53,auto=left,every node/.style={circle,scale=0.6}]

\node[circle, fill=blue] (n31) at (1,6) {};
\node[circle, fill=blue] (n32) at (3,4) {};
\node[circle, fill=blue] (n33) at (1,2) {};
\node[circle, fill=blue] (n34) at (3,0) {};
\node[circle, fill=blue] (n51) at (5,6) {};
\node[circle, fill=blue] (n52) at (7,4) {};
\node[circle, fill=blue] (n53) at (5,2) {};
\node[circle, fill=blue] (n54) at (7,0) {};
\node[circle, fill=red] (n41) at (3,6)  {};
\node[circle, fill=red] (n42) at (1,4)  {};
\node[circle, fill=red] (n43) at (3,2)  {};
\node[circle, fill=red] (n44) at (1,0)  {};
\node[circle, fill=red] (n61) at (7,6)  {};
\node[circle, fill=red] (n62) at (5,4)  {};
\node[circle, fill=red] (n63) at (7,2)  {};
\node[circle, fill=red] (n64) at (5,0)  {};

 \draw[very thick, dashed ] (1,6) -- (1,7);
  \draw[very thick, dashed ] (5,6) -- (5,7);
 \draw[very thick, dashed ] (3,6) -- (3,7);
  \draw[very thick, dashed ] (7,6) -- (7,7);

 \draw[very thick, dashed ] (1,0) -- (1,-1);
  \draw[very thick, dashed ] (3,0) -- (3,-1);
 \draw[very thick, dashed ] (5,0) -- (5,-1);
  \draw[very thick, dashed ] (7,0) -- (7,-1);

\foreach \from/\to in {n31/n41,n41/n51,n51/n61,n32/n42,n42/n32,n32/n62,n52/n62,n33/n43,n43/n53,n53/n63,n31/n42,n42/n33,n41/n32,n32/n43,n51/n62,n62/n53,n61/n52,n52/n63,n54/n64,n44/n34,n34/n64,n34/n43,n54/n63,n33/n44,n53/n64}
  \draw[line width=1.5pt] (\from) -- (\to);

\node[circle, fill=blue] (n11) at (9,7) {};
\node[circle, fill=blue] (n12) at (9,6) {};
\node[circle, fill=blue] (n13) at (9,5) {};
\node[circle, fill=blue] (n14) at (9,4) {};
\node[circle, fill=blue] (n15) at (9,3) {};
\node[circle, fill=blue] (n16) at (9,2) {};
\node[circle, fill=blue] (n17) at (9,1) {};
\node[circle, fill=blue] (n18) at (9,0) {};
\node[circle, fill=red] (n21) at (12,7)  {};
\node[circle, fill=red] (n22) at (12,6)  {};
\node[circle, fill=red] (n23) at (12,5)  {};
\node[circle, fill=red] (n24) at (12,4)  {};
\node[circle, fill=red] (n25) at (12,3)  {};
\node[circle, fill=red] (n26) at (12,2)  {};
\node[circle, fill=red] (n27) at (12,1)  {};
\node[circle, fill=red] (n28) at (12,0)  {};

\foreach \from/\to in {n11/n22,n11/n23,n12/n22,n12/n21,n12/n24,n13/n21,n13/n24,n13/n25,n14/n22,n14/n23,n14/n24,n14/n26,n15/n23,n15/n26,n16/n24,n16/n25,n16/n26,n18/n26,n18/n27,n18/n28,n16/n28,n17/n25,n17/n28,n15/n27,n11/n27,n17/n21,n12/n28,n18/n22}
 \draw[line width=1.5pt] (\from) -- (\to);

\node[circle, fill=teal] (p11) at (14,7) {};
\node[circle, fill=teal] (p12) at (14,6) {};
\node[circle, fill=teal] (p13) at (14,5) {};
\node[circle, fill=teal] (p14) at (14,4) {};
\node[circle, fill=teal] (p15) at (14,3) {};
\node[circle, fill=teal] (p16) at (14,2) {};
\node[circle, fill=teal] (p17) at (14,1) {};
\node[circle, fill=teal] (p18) at (14,0) {};
\node[circle, fill=purple] (p21) at (17,7)  {};
\node[circle, fill=purple] (p22) at (17,6)  {};
\node[circle, fill=purple] (p23) at (17,5)  {};
\node[circle, fill=purple] (p24) at (17,4)  {};
\node[circle, fill=purple] (p25) at (17,3)  {};
\node[circle, fill=purple] (p26) at (17,2)  {};
\node[circle, fill=purple] (p27) at (17,1)  {};
\node[circle, fill=purple] (p28) at (17,0)  {};

\foreach \from/\to in {p11/p21,p12/p22,p13/p23,p14/p24,p15/p25,p16/p26,p17/p27,p18/p28}
 \draw[cyan, line width=1.5pt] (\from) -- (\to);

\draw (4,-1.5)     node[scale=1.8]    {(a)};
\draw (10.5,-1.5)  node[scale=1.8]  {(b)};
\draw (15.5,-1.5)  node[scale=1.8]   {(c)};

\end{tikzpicture}

\caption{Schematic representation of the hopping terms in the Hamiltonian (\ref{ham}) for cylindrical boundary conditions. (a) Example using the $4\times 4$ square lattice, with sites coloured red and blue to highlight the bipartite structure. Compared to the previous Fig. \ref{fig1}, there are four additional edges connecting the top and bottom of the lattice. (b) The equivalent bipartite graph representation that leads to Eq. (\ref{bmatrixcbc}) also has four additional edges compared to the corresponding graph of Fig. \ref{fig1}. (c) Transformed bipartite graph representation resulting from the diagonalisation of $B$. The different colours used, compared to the previous figure, are only to emphasise that a different basis transformation is required.}

\label{fig2}
\end{figure}

For (\ref{bmatrixcbc}) the eigenvalues are found to be given by the multiset 
\begin{align}
{\mathcal S}^4_{{\rm c}}&=\left\{\frac{-3-\sqrt{5}}{2},\frac{1-\sqrt{5}}{2},
\frac{1-\sqrt{5}}{2},\frac{-3+\sqrt{5}}{2},\right. \nonumber \\
&\qquad\qquad\quad\quad\left.\frac{5-\sqrt{5}}{2},\frac{1+\sqrt{5}}{2},\frac{1+\sqrt{5}}{2},\frac{5+\sqrt{5}}{2}\right\}.
\label{setcbc}
\end{align}
For the case of a general $L\times L$ lattice with cylindrical boundary conditions, the eigenvalues of the adjacency matrix  $A$ can again be directly obtained via Fourier series - see Appendix. The eigenvalues of $A$have the form
\begin{align}
2\cos\left(\frac{2\pi j}{L}\right)+2\cos\left(\frac{\pi k}{L+1}\right) \qquad  
j,k\in\{1,\dots,L\} .
\label{formulacbc}
\end{align}
Here, the magnitudes of the elements in (\ref{setcbc}) precisely correspond to the positive values arising from 
(\ref{formulacbc}) for $L=4$. 

\subsection{Toroidal boundary conditions}

For {\it toroidal} boundary conditions, additional non-zero entries arise in  the adjacency matrix. A pictorial representation of the $4\times 4$ lattice is provided in Fig. \ref{fig3} (a). The adjacency submatrix $B$ may be read off from the graph in Fig. \ref{fig3} (b), which is topologically equivalent to Fig. \ref{fig3} (a). It reads  
\begin{align}
B=
\left(
\begin{array}{cccccccc}
{\mathbf 1} & 1 & 1 & 0 & 0 & 0 & {\mathbf 1} & 0 \\
1 & 1 & 0 & 1 & 0 & 0 & 0 & {\mathbf 1} \\
1 & 0 & {\mathbf 1} & 1 & 1 & 0 & 0 & 0 \\
0 & 1 & 1 & 1 & 0 & 1 & 0 & 0 \\
0 & 0 & 1 & 0 & {\mathbf 1} & 1 & 1 & 0 \\
0 & 0 & 0 & 1 & 1 & 1 & 0 & 1 \\
{\mathbf 1} & 0 & 0 & 0 & 1 & 0 & {\mathbf 1} & 1 \\
0 & {\mathbf 1} & 0 & 0 & 0 & 1 & 1 & 1 
\end{array}
\right) 
\label{bmatrixtbc}
\end{align}
Note that 
Fig. \ref{fig3} (b) differs from Fig. \ref{fig1} (b) by the inclusion of 8 additional edges. This leads to eight additional non-zero entries in $B$, depicted by boldface in Eq. (\ref{bmatrixtbc}). 
Each row and column of $B$ has four non-zero entries, reflecting the property that each site has four nearest neighbours once toroidal boundary conditions are imposed.

\begin{figure}[h]
\centering
\captionsetup{width=0.85\linewidth}
\begin{tikzpicture}[scale=.53,auto=left,every node/.style={circle,scale=0.6}]

\node[circle, fill=blue] (n31) at (1,6) {};
\node[circle, fill=blue] (n32) at (3,4) {};
\node[circle, fill=blue] (n33) at (1,2) {};
\node[circle, fill=blue] (n34) at (3,0) {};
\node[circle, fill=blue] (n51) at (5,6) {};
\node[circle, fill=blue] (n52) at (7,4) {};
\node[circle, fill=blue] (n53) at (5,2) {};
\node[circle, fill=blue] (n54) at (7,0) {};
\node[circle, fill=red] (n41) at (3,6)  {};
\node[circle, fill=red] (n42) at (1,4)  {};
\node[circle, fill=red] (n43) at (3,2)  {};
\node[circle, fill=red] (n44) at (1,0)  {};
\node[circle, fill=red] (n61) at (7,6)  {};
\node[circle, fill=red] (n62) at (5,4)  {};
\node[circle, fill=red] (n63) at (7,2)  {};
\node[circle, fill=red] (n64) at (5,0)  {};

 \draw[very thick, dashed ] (1,6) -- (1,7);
  \draw[very thick, dashed ] (5,6) -- (5,7);
 \draw[very thick, dashed ] (3,6) -- (3,7);
  \draw[very thick, dashed ] (7,6) -- (7,7);  
  
 \draw[very thick, dashed ] (7,0) -- (8,0);
  \draw[very thick, dashed ] (7,2) -- (8,2);
 \draw[very thick, dashed ] (7,4) -- (8,4);
  \draw[very thick, dashed ] (7,6) -- (8,6);  

 \draw[very thick, dashed ] (1,0) -- (1,-1);
  \draw[very thick, dashed ] (3,0) -- (3,-1);
 \draw[very thick, dashed ] (5,0) -- (5,-1);
  \draw[very thick, dashed ] (7,0) -- (7,-1);
  
 \draw[very thick, dashed ] (1,6) -- (0,6);
  \draw[very thick, dashed ] (1,4) -- (0,4);
 \draw[very thick, dashed ] (1,2) -- (0,2);
  \draw[very thick, dashed ] (1,0) -- (0,0);

\foreach \from/\to in {n31/n41,n41/n51,n51/n61,n32/n42,n42/n32,n32/n62,n52/n62,n33/n43,n43/n53,n53/n63,n31/n42,n42/n33,n41/n32,n32/n43,n51/n62,n62/n53,n61/n52,n52/n63,n54/n64,n44/n34,n34/n64,n34/n43,n54/n63,n33/n44,n53/n64}
  \draw[line width=1.5pt] (\from) -- (\to);

\node[circle, fill=blue] (n11) at (9,7) {};
\node[circle, fill=blue] (n12) at (9,6) {};
\node[circle, fill=blue] (n13) at (9,5) {};
\node[circle, fill=blue] (n14) at (9,4) {};
\node[circle, fill=blue] (n15) at (9,3) {};
\node[circle, fill=blue] (n16) at (9,2) {};
\node[circle, fill=blue] (n17) at (9,1) {};
\node[circle, fill=blue] (n18) at (9,0) {};
\node[circle, fill=red] (n21) at (12,7)  {};
\node[circle, fill=red] (n22) at (12,6)  {};
\node[circle, fill=red] (n23) at (12,5)  {};
\node[circle, fill=red] (n24) at (12,4)  {};
\node[circle, fill=red] (n25) at (12,3)  {};
\node[circle, fill=red] (n26) at (12,2)  {};
\node[circle, fill=red] (n27) at (12,1)  {};
\node[circle, fill=red] (n28) at (12,0)  {};

\foreach \from/\to in {n11/n22,n11/n23,n12/n22,n12/n21,n12/n24,n13/n21,n13/n24,n13/n25,n14/n22,n14/n23,n14/n24,n14/n26,n15/n23,n15/n26,n16/n24,n16/n25,n16/n26,n18/n26,n18/n27,n18/n28,n16/n28,n17/n25,n17/n28,n15/n27,n11/n21,n13/n23,n15/n25,n17/n27,n11/n27,n17/n21,n12/n28,n18/n22}
 \draw[line width=1.5pt] (\from) -- (\to);

\node[circle, fill=teal] (p11) at (14,7) {};
\node[circle, fill=teal] (p12) at (14,6) {};
\node[circle, fill=teal] (p13) at (14,5) {};
\node[circle, fill=teal] (p14) at (14,4) {};
\node[circle, fill=teal] (p15) at (14,3) {};
\node[circle, fill=teal] (p16) at (14,2) {};
\node[circle, fill=teal] (p17) at (14,1) {};
\node[circle, fill=teal] (p18) at (14,0) {};
\node[circle, fill=purple] (p21) at (17,7)  {};
\node[circle, fill=purple] (p22) at (17,6)  {};
\node[circle, fill=purple] (p23) at (17,5)  {};
\node[circle, fill=purple] (p24) at (17,4)  {};
\node[circle, fill=purple] (p25) at (17,3)  {};
\node[circle, fill=purple] (p26) at (17,2)  {};
\node[circle, fill=purple] (p27) at (17,1)  {};
\node[circle, fill=purple] (p28) at (17,0)  {};

\foreach \from/\to in {p11/p21,p12/p22,p13/p23,p14/p24,p15/p25,p16/p26,p17/p27,p18/p28}
 \draw[green, line width=1.5pt] (\from) -- (\to);

\draw (4,-1.5)     node[scale=1.8]    {(a)};
\draw (10.5,-1.5)  node[scale=1.8]  {(b)};
\draw (15.5,-1.5)  node[scale=1.8]   {(c)};

\end{tikzpicture}

\caption{Schematic representation of the hopping terms in the Hamiltonian (\ref{ham}) for toroidal boundary conditions. (a) Example using the $4\times 4$ square lattice, with sites coloured red and blue to highlight the bipartite structure. (b) Equivalent bipartite graph representation that allows the adjacency submatrix $B$ to be determined, as given by Eq. (\ref{bmatrixtbc}). All vertices have four edges, so each row and column of (\ref{bmatrixtbc}) has four entries, reflecting the horizontal and vertical translational invariance of the lattice.   (c) Transformed bipartite graph representation resulting from the diagonalisation of $B$. As before, the different colours used are only to emphasise that the basis transformation is different from those required for the other boundary conditions.}

\label{fig3}
\end{figure}
For ({\ref{bmatrixtbc}) the eigenvalues are found to be given by the multiset 
\begin{align}
{\mathcal S}^4_{{\rm t}}=\{-2,0,0,0,2,2,2,4\}.
\label{settbc}
\end{align}
For the case of a general $L\times L$ lattice with toroidal boundary conditions, the eigenvalues of the adjacency matrix  $A$ can be directly obtained via Fourier series - see Appendix. The eigenvalues of $A$ have the form
\begin{align}
2\cos\left(\frac{2\pi j}{L}\right)+2\cos\left(\frac{2\pi k}{L}\right) \qquad  
j,k\in\{1,\dots,L\} .
\label{formulatbc}
\end{align}
It is readily verified for $L=4$ that Eq. (\ref{formulatbc}) provides the values
\begin{align*}
\{-4,-2,-2,-2,-2,0,0,0,0,0,0,2,2,2,2,4\}, 
\end{align*}
representing the union ${\mathcal S}^t_4 \cup (-{\mathcal S}_4^t)$  from ({\ref{settbc}), as expected.

\subsection{Ground-state energy and imbalance fluctuations}
Having set up the framework for the three different types of boundary conditions in the previous subsections, this is now employed to investigate some ground state properties. The convention is adopted to label the eigenvalues $\varepsilon_j$ of $B$ such that $j=1$ corresponds to the maximum  $|\varepsilon_j|$ in each set. That is  
\begin{align}
|\varepsilon_1|=
\begin{cases}
\displaystyle 4\cos\left(\frac{\pi}{L+1}\right),  &  {\rm open}, \\
\displaystyle 2+ 2\cos\left(\frac{\pi}{L+1}\right), & {\rm cylindrical}, \\
\displaystyle 4, & {\rm toroidal}.
\end{cases}
\label{high}
\end{align}

Next some analytic solutions of the Bethe Ansatz eqs. (\ref{bae}) are presented, starting with $N=2$. For arbitrary $m$, equivalently arbitrary $L$, choosing the quantum numbers $N_1=2$, and $N_j=0$ otherwise, leads to the coupled Bethe Ansatz equations   
\begin{align}
 \frac{2v_1}{v_1-v_2}
+\frac{(v_1-\varepsilon_1^2)^{2}}{ 16U^2(v_1-v_2)}
&=1+ \frac{2\varepsilon_1^2}{v_1-\varepsilon_1^2}, \label{bae2a}\\
 \frac{2v_2}{v_2-v_1}
+\frac{(v_2-\varepsilon_1^2)^{2}}{ 16U^2 (v_2-v_1)}
&=1+\frac{2\varepsilon_1^2}{v_2-\varepsilon_1^2}, \label{bae2b}
\end{align}
with corresponding energy eigenvalue
\begin{align}
E=4U +4U   \frac{2 \varepsilon_1^2}{v_1-\varepsilon_1^2}
+4U   \frac{2 \varepsilon_1^2}{v_2-\varepsilon_1^2}.
\label{nrg2}
\end{align}
Setting 
\begin{align*}
\phi&= -2U - 2\sqrt{U^2+\varepsilon_1^2}
\end{align*}
it is straightforward to verify that 
\begin{align*}
v_1&=\varepsilon_1^2+2U\phi + 2U\sqrt{\phi^2 -8 \varepsilon_1^2 }, \\
v_2&=\varepsilon_1^2+2U\phi - 2U\sqrt{\phi^2 -8 \varepsilon_1^2 }
\end{align*}
satisfy (\ref{bae2a},\ref{bae2b}), while substituting this solution into (\ref{nrg2}) yields
\begin{align}
E = 2U - 2\sqrt{U^2+\varepsilon_1^2} .
\label{nrg2gs}
\end{align}

Moving ahead to $N=3$, set  $N_1=3$ and $N_j=0$ otherwise. This leads to the Bethe Ansatz equations 
\begin{align}
\sum_{p\neq n}^3 \frac{2v_n}{v_n-v_p}
+\frac{\displaystyle (v_n-\varepsilon_1^2)^{3}}{\displaystyle 16U^2\prod_{p\neq n}^3(v_n-v_p)}
&=2+ \frac{3\varepsilon_1^2}{v_n-\varepsilon_1^2}, \qquad n=1,2,3. 
\label{bae3}
\end{align}
Let $w_1,\,w_2,\,w_3$ denote the roots of the polynomial 
\begin{align}
p(\lambda)=|\varepsilon_1| \lambda^3- \sigma \lambda^2 +2\sigma \lambda -6|\varepsilon_1|
\label{cubic}
\end{align}
where
\begin{align*}
\sigma =  -4U-|\varepsilon_1| - 2\sqrt{4U^2  +2U|\varepsilon_1| +\varepsilon_1^2}.
\end{align*}
Setting $v_n=\varepsilon_1^2+4U|\varepsilon_1| w_n$ provides a solution to (\ref{bae3}), while substitution into
\begin{align*}
E=9U +4U   \sum_{n=1}^3  \frac{3 \varepsilon_1^2}{v_n-\varepsilon_1^2}
\end{align*}
yields
\begin{align}
E=5U-|\varepsilon_1|-2\sqrt{ 4U^2 +2U|\varepsilon_1|+\varepsilon_1^2}.
\label{nrg3}
\end{align}

Note that the $L$ dependence in (\ref{nrg2}) and (\ref{nrg3}) only enters through the variable 
$\varepsilon_1$. In the case of $4\times 4$ square lattices, numerical calculations indicate that the expressions (\ref{nrg2}) and (\ref{nrg3}) provide the $N=2$ and $N=3$ ground-state energies respectively, for all three types of boundary conditions discussed previously. These expressions also hold for all values of 
$L$ in some neighbourhood of $U=0$, since they produce the correct ground-state energy values in the limit 
$U\rightarrow 0$.  

Define the {\it imbalance fluctuation} ${\mathcal I}$ through
\begin{align*}
{\mathcal I} =\langle (\hat{N}_a-\hat{N}_b)^2\rangle - \langle \hat{N}_a-\hat{N}_b\rangle^2 
\end{align*}
For any eigenstate that is non-degenerate, $\langle \hat{N}_a-\hat{N}_b\rangle=0$ by the interchange symmetry $a_j \leftrightarrow b_j$. Next, use of the Hellmann-Feynman theorem produces 
\begin{align*}
\langle (\hat{N}_a-\hat{N}_b)^2\rangle=\frac{\partial E}{\partial U} ,
\end{align*}
where $E$ is the corresponding eigenvalue. This leads to the ground-state imbalance fluctuation  
\begin{align}
{\mathcal I} = 2- \frac{2U}{\sqrt{U^2+\varepsilon_1^2}}
\label{i2}
\end{align}
through (\ref{nrg2}) for $N=2$, and 
\begin{align}
{\mathcal I} = 5- \frac{8U+2|\varepsilon_1|}{\sqrt{4U^2+2U|\varepsilon_1|+\varepsilon_1^2}}
\label{i3}
\end{align}
through (\ref{nrg3}) for $N=3$. Note that $\displaystyle \lim_{U\rightarrow \infty} {\mathcal I}$ is zero for $N=2$, and non-zero for $N=3$. This illustrates a parity effect in the properties of the model, dependent on whether the particle number is even or odd.
 
\section{Integrable defect}

The previous examples, dealing with the three different types of boundary conditions, illustrate how the integrability is a structural feature of the class of Hamiltonians of the form (\ref{ham}), and is not reliant on any particular properties of the adjacency submatrix $B$ (other than it is a real, symmetric matrix). This is due to the formulae (\ref{con1},\ref{con2}) for the conserved operators holding independent of any particular properties of $B$. This opens an avenue for investigating integrability preserving defects in the system.
 
Properties of $B$ do, however, strongly influence the character of the Bethe Ansatz solution; the solution depends explicitly on the eigenvalues of $B$. For the previous three classes of boundary conditions it was found that the Bethe Ansatz solution can be expressed in closed form, because the eigenvalues are obtained in terms of trigonometric functions. It will be seen that the closed form property is no longer present for defective integrable models. This will be illustrated through an example, where the defect is associated with the boundary of an open system for $L=8$.

\begin{figure}[h]
\centering
\captionsetup{width=.85\linewidth}
\begin{tikzpicture}[scale=.4,auto=left,every node/.style={circle,scale=0.6}]

\node[circle, fill=blue] (n11) at (-3,6) {};
\node[circle, fill=blue] (n12) at (-1,4) {};
\node[circle, fill=blue] (n13) at (-3,2) {};
\node[circle, fill=blue] (n14) at (-1,0) {};
\node[circle, fill=blue] (n15) at (-3,-2) {};
\node[circle, fill=blue] (n16) at (-1,-4) {};
\node[circle, fill=blue] (n17) at (-3,-6) {};
\node[circle, fill=blue] (n18) at (-1,-8) {};
\node[circle, fill=blue] (n31) at (1,6) {};
\node[circle, fill=blue] (n32) at (3,4) {};
\node[circle, fill=blue] (n33) at (1,2) {};
\node[circle, fill=blue] (n34) at (3,0) {};
\node[circle, fill=blue] (n35) at (1,-2) {};
\node[circle, fill=blue] (n36) at (3,-4) {};
\node[circle, fill=blue] (n37) at (1,-6) {};
\node[circle, fill=blue] (n38) at (3,-8) {};
\node[circle, fill=blue] (n51) at (5,6) {};
\node[circle, fill=blue] (n52) at (7,4) {};
\node[circle, fill=blue] (n53) at (5,2) {};
\node[circle, fill=blue] (n54) at (7,0) {};
\node[circle, fill=blue] (n55) at (5,-2) {};
\node[circle, fill=blue] (n56) at (7,-4) {};
\node[circle, fill=blue] (n57) at (5,-6) {};
\node[circle, fill=blue] (n58) at (7,-8) {};
\node[circle, fill=blue] (n71) at (9,6) {};
\node[circle, fill=blue] (n72) at (11,4) {};
\node[circle, fill=blue] (n73) at (9,2) {};
\node[circle, fill=blue] (n74) at (11,0) {};
\node[circle, fill=blue] (n75) at (9,-2) {};
\node[circle, fill=blue] (n76) at (11,-4) {};
\node[circle, fill=blue] (n77) at (9,-6) {};
\node[circle, fill=blue] (n78) at (11,-8) {};
\node[circle, fill=red] (n21) at (-1,6)  {};
\node[circle, fill=red] (n22) at (-3,4)  {};
\node[circle, fill=red] (n23) at (-1,2)  {};
\node[circle, fill=red] (n24) at (-3,0)  {};
\node[circle, fill=red] (n25) at (-1,-2)  {};
\node[circle, fill=red] (n26) at (-3,-4)  {};
\node[circle, fill=red] (n27) at (-1,-6)  {};
\node[circle, fill=red] (n28) at (-3,-8)  {};

\node[circle, fill=red] (n41) at (3,6)  {};
\node[circle, fill=red] (n42) at (1,4)  {};
\node[circle, fill=red] (n43) at (3,2)  {};
\node[circle, fill=red] (n44) at (1,0)  {};
\node[circle, fill=red] (n45) at (3,-2)  {};
\node[circle, fill=red] (n46) at (1,-4)  {};
\node[circle, fill=red] (n47) at (3,-6)  {};
\node[circle, fill=red] (n48) at (1,-8)  {};

\node[circle, fill=red] (n61) at (7,6)  {};
\node[circle, fill=red] (n62) at (5,4)  {};
\node[circle, fill=red] (n63) at (7,2)  {};
\node[circle, fill=red] (n64) at (5,0)  {};
\node[circle, fill=red] (n65) at (7,-2)  {};
\node[circle, fill=red] (n66) at (5,-4)  {};
\node[circle, fill=red] (n67) at (7,-6)  {};
\node[circle, fill=red] (n68) at (5,-8)  {};

\node[circle, fill=red] (n81) at (11,6)  {};
\node[circle, fill=red] (n82) at (9,4)  {};
\node[circle, fill=red] (n83) at (11,2)  {};
\node[circle, fill=red] (n84) at (9,0)  {};
\node[circle, fill=red] (n85) at (11,-2)  {};
\node[circle, fill=red] (n86) at (9,-4)  {};
\node[circle, fill=red] (n87) at (11,-6)  {};
\node[circle, fill=red] (n88) at (9,-8)  {};

\node[circle, fill=blue] (p11) at (14,6) {};
\node[circle, fill=blue] (p12) at (16,4) {};
\node[circle, fill=blue] (p13) at (14,2) {};
\node[circle, fill=blue] (p14) at (16,0) {};
\node[circle, fill=blue] (p15) at (14,-2) {};
\node[circle, fill=blue] (p16) at (16,-4) {};
\node[circle, fill=blue] (p17) at (14,-6) {};
\node[circle, fill=blue] (p18) at (16,-8) {};
\node[circle, fill=blue] (p31) at (18,6) {};
\node[circle, fill=blue] (p32) at (20,4) {};
\node[circle, fill=blue] (p33) at (18,2) {};
\node[circle, fill=blue] (p34) at (20,0) {};
\node[circle, fill=blue] (p35) at (18,-2) {};
\node[circle, fill=blue] (p36) at (20,-4) {};
\node[circle, fill=blue] (p37) at (18,-6) {};
\node[circle, fill=blue] (p38) at (20,-8) {};
\node[circle, fill=blue] (p51) at (22,6) {};
\node[circle, fill=blue] (p52) at (24,4) {};
\node[circle, fill=blue] (p53) at (22,2) {};
\node[circle, fill=blue] (p54) at (24,0) {};
\node[circle, fill=blue] (p55) at (22,-2) {};
\node[circle, fill=blue] (p56) at (24,-4) {};
\node[circle, fill=blue] (p57) at (22,-6) {};
\node[circle, fill=blue] (p58) at (24,-8) {};
\node[circle, fill=blue] (p71) at (26,6) {};
\node[circle, fill=blue] (p72) at (28,4) {};
\node[circle, fill=blue] (p73) at (26,2) {};
\node[circle, fill=blue] (p74) at (28,0) {};
\node[circle, fill=blue] (p75) at (26,-2) {};
\node[circle, fill=blue] (p76) at (28,-4) {};
\node[circle, fill=blue] (p77) at (26,-6) {};
\node[circle, fill=blue] (p78) at (28,-8) {};

\node[circle, fill=red] (p21) at (16,6)  {};
\node[circle, fill=red] (p22) at (14,4)  {};
\node[circle, fill=red] (p23) at (16,2)  {};
\node[circle, fill=red] (p24) at (14,0)  {};
\node[circle, fill=red] (p25) at (16,-2)  {};
\node[circle, fill=red] (p26) at (14,-4)  {};
\node[circle, fill=red] (p27) at (16,-6)  {};
\node[circle, fill=red] (p28) at (14,-8)  {};

\node[circle, fill=red] (p41) at (20,6)  {};
\node[circle, fill=red] (p42) at (18,4)  {};
\node[circle, fill=red] (p43) at (20,2)  {};
\node[circle, fill=red] (p44) at (18,0)  {};
\node[circle, fill=red] (p45) at (20,-2)  {};
\node[circle, fill=red] (p46) at (18,-4)  {};
\node[circle, fill=red] (p47) at (20,-6)  {};
\node[circle, fill=red] (p48) at (18,-8)  {};

\node[circle, fill=red] (p61) at (24,6)  {};
\node[circle, fill=red] (p62) at (22,4)  {};
\node[circle, fill=red] (p63) at (24,2)  {};
\node[circle, fill=red] (p64) at (22,0)  {};
\node[circle, fill=red] (p65) at (24,-2)  {};
\node[circle, fill=red] (p66) at (22,-4)  {};
\node[circle, fill=red] (p67) at (24,-6)  {};
\node[circle, fill=red] (p68) at (22,-8)  {};

\node[circle, fill=red] (p81) at (28,6)  {};
\node[circle, fill=red] (p82) at (26,4)  {};
\node[circle, fill=red] (p83) at (28,2)  {};
\node[circle, fill=red] (p84) at (26,0)  {};
\node[circle, fill=red] (p85) at (28,-2)  {};
\node[circle, fill=red] (p86) at (26,-4)  {};
\node[circle, fill=red] (p87) at (28,-6)  {};
\node[circle, fill=red] (p88) at (26,-8)  {};

\foreach \from/\to in {n41/n51,n31/n41,n51/n61,n32/n42,n32/n62,n52/n62,n33/n43,n43/n53,n53/n63,n31/n42,n42/n33,n41/n32,n32/n43,n51/n62,n62/n53,n61/n52,n52/n63,n54/n64,n44/n34,n34/n64,n34/n43,n54/n63,n33/n44,n53/n64}
  \draw[line width=1.5pt] (\from) -- (\to);

\foreach \from/\to in {n45/n55,n35/n45,n55/n65,n36/n46,n36/n66,n56/n66,n37/n47,n47/n57,n57/n67,n35/n46,n46/n37,n45/n36,n36/n47,n55/n66,n66/n57,n65/n56,n56/n67,n58/n68,n48/n38,n38/n68,n38/n47,n58/n67,n37/n48,n57/n68}
  \draw[line width=1.5pt] (\from) -- (\to);	
	
\foreach \from/\to in {n11/n21,n21/n31,n61/n71,n71/n81,n12/n22,n22/n42,n52/n82,n72/n82,n13/n23,n23/n33,n63/n73,n73/n83,n14/n24,n24/n44,n54/n84,n74/n84,n15/n25,n25/n35,n65/n75,n75/n85,n16/n26,n16/n46,n56/n86,n76/n86,n17/n27,n27/n37,n67/n77,n77/n87,n18/n28,n18/n48,n58/n88,n78/n88}
  \draw[line width=1.5pt] (\from) -- (\to);		

\foreach \from/\to in {n11/n22,n22/n13,n13/n24,n24/n15,n15/n26,n26/n17,n17/n28,n21/n12,n12/n23,n23/n14,n14/n25,n25/n16,n16/n27,n27/n18,n44/n35,n34/n45}
\draw[line width=1.5pt] (\from) -- (\to);		

\foreach \from/\to in {n71/n82,n82/n73,n73/n84,n84/n75,n75/n86,n86/n77,n77/n88,n81/n72,n72/n83,n83/n74,n74/n85,n85/n76,n76/n87,n87/n78,n64/n55,n54/n65}
\draw[line width=1.5pt] (\from) -- (\to);


\foreach \from/\to in {p31/p41,p51/p61,p32/p42,p32/p62,p52/p62,p33/p43,p43/p53,p53/p63,p31/p42,p42/p33,p41/p32,p32/p43,p51/p62,p62/p53,p61/p52,p52/p63,p54/p64,p44/p34,p34/p64,p34/p43,p54/p63,p33/p44,p53/p64}
  \draw[line width=1.5pt] (\from) -- (\to);

\foreach \from/\to in {p45/p55,p35/p45,p55/p65,p36/p46,p36/p66,p56/p66,p37/p47,p47/p57,p57/p67,p35/p46,p46/p37,p45/p36,p36/p47,p55/p66,p66/p57,p65/p56,p56/p67,p58/p68,p48/p38,p38/p68,p38/p47,p58/p67,p37/p48,p57/p68}
  \draw[line width=1.5pt] (\from) -- (\to);

\foreach \from/\to in {p11/p21,p21/p31,
p61/p71,p71/p81,
p12/p22,p12/p42,
p52/p82,p72/p82,
p13/p23,p23/p33,
p63/p73,p73/p83,
p14/p24,p14/p44,
p54/p84,p74/p84,
p15/p25,p25/p35,
p65/p75,p75/p85,
p16/p26,p16/p46,
p56/p86,p76/p86,
p17/p27,p27/p37,
p67/p77,p77/p87,
p18/p28,p18/p48,
p58/p88,p78/p88}
  \draw[line width=1.5pt] (\from) -- (\to);

\foreach \from/\to in {p11/p22,p22/p13,p13/p24,p24/p15,p15/p26,p26/p17,p17/p28,p21/p12,p12/p23,p23/p14,p14/p25,p25/p16,p16/p27,p27/p18,p44/p35,p34/p45}
\draw[line width=1.5pt] (\from) -- (\to);		

\foreach \from/\to in {p71/p82,p82/p73,p73/p84,p84/p75,p75/p86,p86/p77,p77/p88,p81/p72,p72/p83,p83/p74,p74/p85,p85/p76,p76/p87,p87/p78,p64/p55,p54/p65}
\draw[line width=1.5pt] (\from) -- (\to);


\draw (4,-10)     node[scale=1.8]    {(a)};
\draw (21,-10)  node[scale=1.8]  {(b)};

\end{tikzpicture}

\caption{Schematic representation of the square lattice with a boundary defect. (a) An example using the $8\times 8$ square lattice, with open boundary conditions. (b) Illustration of a boundary defect through removal of an edge of the graph from the top boundary.}

\label{fig4}
\end{figure} 

Consider Fig. \ref{fig4} (a) depicting an $8\times 8$ open lattice, and Fig. \ref{fig4} (b) with a defect on the top boundary. 
The adjacency submatrix $B$ for case (b) may be read off from the graph, as in the previous examples; 
\begin{align}
B=
\left(
\begin{array}{c} 
0\ \, 0\ \, 0\ \, 1\ \, 1\ \, 0\ \, 0\ \, 0\ \, 0\ \, 0\ \, 0\ \, 0\ \, 0\ \, 0\ \, 0\ \, 0\ \, 0\ \, 0\ \, 0\ \, 0\ \, 0\ \, 0\ \, 0\ \, 0\ \, 0\ \, 0\ \, 0\ \, 0\ \, 0\ \, 0\ \, 0\ \, 0
\\ 
0\ \, 0\ \, 1\ \, 1\ \, 0\ \, 1\ \, 0\ \, 0\ \, 0\ \, 0\ \, 0\ \, 0\ \, 0\ \, 0\ \, 0\ \, 0\ \, 0\ \, 0\ \, 0\ \, 0\ \, 0\ \, 0\ \, 0\ \, 0\ \, 0\ \, 0\ \, 0\ \, 0\ \, 0\ \, 0\ \, 0\ \, 0
\\ 
0\ \, 1\ \, {\mathbf 0}\ \, 0\ \, 0\ \, 0\ \, 1\ \, 0\ \, 0\ \, 0\ \, 0\ \, 0\ \, 0\ \, 0\ \, 0\ \, 0\ \, 0\ \, 0\ \, 0\ \, 0\ \, 0\ \, 0\ \, 0\ \, 0\ \, 0\ \, 0\ \, 0\ \, 0\ \, 0\ \, 0\ \, 0\ \, 0
\\ 
1\ \, 1\ \, 0\ \, 0\ \, 0\ \, 0\ \, 0\ \, 1\ \, 0\ \, 0\ \, 0\ \, 0\ \, 0\ \, 0\ \, 0\ \, 0\ \, 0\ \, 0\ \, 0\ \, 0\ \, 0\ \, 0\ \, 0\ \, 0\ \, 0\ \, 0\ \, 0\ \, 0\ \, 0\ \, 0\ \, 0\ \, 0
\\
1\ \, 0\ \, 0\ \, 0\ \, 0\ \, 0\ \, 0\ \, 1\ \, 1\ \, 0\ \, 0\ \, 0\ \, 0\ \, 0\ \, 0\ \, 0\ \, 0\ \, 0\ \, 0\ \, 0\ \, 0\ \, 0\ \, 0\ \, 0\ \, 0\ \, 0\ \, 0\ \, 0\ \, 0\ \, 0\ \, 0\ \, 0
\\
0\ \, 1\ \, 0\ \, 0\ \, 0\ \, 0\ \, 1\ \, 1\ \, 0\ \, 1\ \, 0\ \, 0\ \, 0\ \, 0\ \, 0\ \, 0\ \, 0\ \, 0\ \, 0\ \, 0\ \, 0\ \, 0\ \, 0\ \, 0\ \, 0\ \, 0\ \, 0\ \, 0\ \, 0\ \, 0\ \, 0\ \, 0
\\ 
0\ \, 0\ \, 1\ \, 0\ \, 0\ \, 1\ \, 1\ \, 0\ \, 0\ \, 0\ \, 1\ \, 0\ \, 0\ \, 0\ \, 0\ \, 0\ \, 0\ \, 0\ \, 0\ \, 0\ \, 0\ \, 0\ \, 0\ \, 0\ \, 0\ \, 0\ \, 0\ \, 0\ \, 0\ \, 0\ \, 0\ \, 0
\\
0\ \, 0\ \, 0\ \, 1\ \, 1\ \, 1\ \, 0\ \, 0\ \, 0\ \, 0\ \, 0\ \, 1\ \, 0\ \, 0\ \, 0\ \, 0\ \, 0\ \, 0\ \, 0\ \, 0\ \, 0\ \, 0\ \, 0\ \, 0\ \, 0\ \, 0\ \, 0\ \, 0\ \, 0\ \, 0\ \, 0\ \, 0
\\
0\ \, 0\ \, 0\ \, 0\ \, 1\ \, 0\ \, 0\ \, 0\ \, 0\ \, 0\ \, 0\ \, 1\ \, 1\ \, 0\ \, 0\ \, 0\ \, 0\ \, 0\ \, 0\ \, 0\ \, 0\ \, 0\ \, 0\ \, 0\ \, 0\ \, 0\ \, 0\ \, 0\ \, 0\ \, 0\ \, 0\ \, 0
\\
0\ \, 0\ \, 0\ \, 0\ \, 0\ \, 1\ \, 0\ \, 0\ \, 0\ \, 0\ \, 1\ \, 1\ \, 0\ \, 1\ \, 0\ \, 0\ \, 0\ \, 0\ \, 0\ \, 0\ \, 0\ \, 0\ \, 0\ \, 0\ \, 0\ \, 0\ \, 0\ \, 0\ \, 0\ \, 0\ \, 0\ \, 0
\\
0\ \, 0\ \, 0\ \, 0\ \, 0\ \, 0\ \, 1\ \, 0\ \, 0\ \, 1\ \, 1\ \, 0\ \, 0\ \, 0\ \, 1\ \, 0\ \, 0\ \, 0\ \, 0\ \, 0\ \, 0\ \, 0\ \, 0\ \, 0\ \, 0\ \, 0\ \, 0\ \, 0\ \, 0\ \, 0\ \, 0\ \, 0
\\
0\ \, 0\ \, 0\ \, 0\ \, 0\ \, 0\ \, 0\ \, 1\ \, 1\ \, 1\ \, 0\ \, 0\ \, 0\ \, 0\ \, 0\ \, 1\ \, 0\ \, 0\ \, 0\ \, 0\ \, 0\ \, 0\ \, 0\ \, 0\ \, 0\ \, 0\ \, 0\ \, 0\ \, 0\ \, 0\ \, 0\ \, 0
\\
0\ \, 0\ \, 0\ \, 0\ \, 0\ \, 0\ \, 0\ \, 0\ \, 1\ \, 0\ \, 0\ \, 0\ \, 0\ \, 0\ \, 0\ \, 1\ \, 1\ \, 0\ \, 0\ \, 0\ \, 0\ \, 0\ \, 0\ \, 0\ \, 0\ \, 0\ \, 0\ \, 0\ \, 0\ \, 0\ \, 0\ \, 0
\\
0\ \, 0\ \, 0\ \, 0\ \, 0\ \, 0\ \, 0\ \, 0\ \, 0\ \, 1\ \, 0\ \, 0\ \, 0\ \, 0\ \, 1\ \, 1\ \, 0\ \, 1\ \, 0\ \, 0\ \, 0\ \, 0\ \, 0\ \, 0\ \, 0\ \, 0\ \, 0\ \, 0\ \, 0\ \, 0\ \, 0\ \, 0
\\
0\ \, 0\ \, 0\ \, 0\ \, 0\ \, 0\ \, 0\ \, 0\ \, 0\ \, 0\ \, 1\ \, 0\ \, 0\ \, 1\ \, 1\ \, 0\ \, 0\ \, 0\ \, 1\ \, 0\ \, 0\ \, 0\ \, 0\ \, 0\ \, 0\ \, 0\ \, 0\ \, 0\ \, 0\ \, 0\ \, 0\ \, 0
\\
0\ \, 0\ \, 0\ \, 0\ \, 0\ \, 0\ \, 0\ \, 0\ \, 0\ \, 0\ \, 0\ \, 1\ \, 1\ \, 1\ \, 0\ \, 0\ \, 0\ \, 0\ \, 0\ \, 1\ \, 0\ \, 0\ \, 0\ \, 0\ \, 0\ \, 0\ \, 0\ \, 0\ \, 0\ \, 0\ \, 0\ \, 0
\\
0\ \, 0\ \, 0\ \, 0\ \, 0\ \, 0\ \, 0\ \, 0\ \, 0\ \, 0\ \, 0\ \, 0\ \, 1\ \, 0\ \, 0\ \, 0\ \, 0\ \, 0\ \, 0\ \, 1\ \, 1\ \, 0\ \, 0\ \, 0\ \, 0\ \, 0\ \, 0\ \, 0\ \, 0\ \, 0\ \, 0\ \, 0
\\
0\ \, 0\ \, 0\ \, 0\ \, 0\ \, 0\ \, 0\ \, 0\ \, 0\ \, 0\ \, 0\ \, 0\ \, 0\ \, 1\ \, 0\ \, 0\ \, 0\ \, 0\ \, 1\ \, 1\ \, 0\ \, 1\ \, 0\ \, 0\ \, 0\ \, 0\ \, 0\ \, 0\ \, 0\ \, 0\ \, 0\ \, 0
\\
0\ \, 0\ \, 0\ \, 0\ \, 0\ \, 0\ \, 0\ \, 0\ \, 0\ \, 0\ \, 0\ \, 0\ \, 0\ \, 0\ \, 1\ \, 0\ \, 0\ \, 1\ \, 1\ \, 0\ \, 0\ \, 0\ \, 1\ \, 0\ \, 0\ \, 0\ \, 0\ \, 0\ \, 0\ \, 0\ \, 0\ \, 0
\\
0\ \, 0\ \, 0\ \, 0\ \, 0\ \, 0\ \, 0\ \, 0\ \, 0\ \, 0\ \, 0\ \, 0\ \, 0\ \, 0\ \, 0\ \, 1\ \, 1\ \, 1\ \, 0\ \, 0\ \, 0\ \, 0\ \, 0\ \, 1\ \, 0\ \, 0\ \, 0\ \, 0\ \, 0\ \, 0\ \, 0\ \, 0
\\
0\ \, 0\ \, 0\ \, 0\ \, 0\ \, 0\ \, 0\ \, 0\ \, 0\ \, 0\ \, 0\ \, 0\ \, 0\ \, 0\ \, 0\ \, 0\ \, 1\ \, 0\ \, 0\ \, 0\ \, 0\ \, 0\ \, 0\ \, 1\ \, 1\ \, 0\ \, 0\ \, 0\ \, 0\ \, 0\ \, 0\ \, 0
\\
0\ \, 0\ \, 0\ \, 0\ \, 0\ \, 0\ \, 0\ \, 0\ \, 0\ \, 0\ \, 0\ \, 0\ \, 0\ \, 0\ \, 0\ \, 0\ \, 0\ \, 1\ \, 0\ \, 0\ \, 0\ \, 0\ \, 1\ \, 1\ \, 0\ \, 1\ \, 0\ \, 0\ \, 0\ \, 0\ \, 0\ \, 0
\\
0\ \, 0\ \, 0\ \, 0\ \, 0\ \, 0\ \, 0\ \, 0\ \, 0\ \, 0\ \, 0\ \, 0\ \, 0\ \, 0\ \, 0\ \, 0\ \, 0\ \, 0\ \, 1\ \, 0\ \, 0\ \, 1\ \, 1\ \, 0\ \, 0\ \, 0\ \, 1\ \, 0\ \, 0\ \, 0\ \, 0\ \, 0
\\
0\ \, 0\ \, 0\ \, 0\ \, 0\ \, 0\ \, 0\ \, 0\ \, 0\ \, 0\ \, 0\ \, 0\ \, 0\ \, 0\ \, 0\ \, 0\ \, 0\ \, 0\ \, 0\ \, 1\ \, 1\ \, 1\ \, 0\ \, 0\ \, 0\ \, 0\ \, 0\ \, 1\ \, 0\ \, 0\ \, 0\ \, 0
\\
0\ \, 0\ \, 0\ \, 0\ \, 0\ \, 0\ \, 0\ \, 0\ \, 0\ \, 0\ \, 0\ \, 0\ \, 0\ \, 0\ \, 0\ \, 0\ \, 0\ \, 0\ \, 0\ \, 0\ \, 1\ \, 0\ \, 0\ \, 0\ \, 0\ \, 0\ \, 0\ \, 1\ \, 1\ \, 0\ \, 0\ \, 0
\\
0\ \, 0\ \, 0\ \, 0\ \, 0\ \, 0\ \, 0\ \, 0\ \, 0\ \, 0\ \, 0\ \, 0\ \, 0\ \, 0\ \, 0\ \, 0\ \, 0\ \, 0\ \, 0\ \, 0\ \, 0\ \, 1\ \, 0\ \, 0\ \, 0\ \, 0\ \, 1\ \, 1\ \, 0\ \, 1\ \, 0\ \, 0
\\
0\ \, 0\ \, 0\ \, 0\ \, 0\ \, 0\ \, 0\ \, 0\ \, 0\ \, 0\ \, 0\ \, 0\ \, 0\ \, 0\ \, 0\ \, 0\ \, 0\ \, 0\ \, 0\ \, 0\ \, 0\ \, 0\ \, 1\ \, 0\ \, 0\ \, 1\ \, 1\ \, 0\ \, 0\ \, 0\ \, 1\ \, 0
\\
0\ \, 0\ \, 0\ \, 0\ \, 0\ \, 0\ \, 0\ \, 0\ \, 0\ \, 0\ \, 0\ \, 0\ \, 0\ \, 0\ \, 0\ \, 0\ \, 0\ \, 0\ \, 0\ \, 0\ \, 0\ \, 0\ \, 0\ \, 1\ \, 1\ \, 1\ \, 0\ \, 0\ \, 0\ \, 0\ \, 0\ \, 1
\\
0\ \, 0\ \, 0\ \, 0\ \, 0\ \, 0\ \, 0\ \, 0\ \, 0\ \, 0\ \, 0\ \, 0\ \, 0\ \, 0\ \, 0\ \, 0\ \, 0\ \, 0\ \, 0\ \, 0\ \, 0\ \, 0\ \, 0\ \, 0\ \, 1\ \, 0\ \, 0\ \, 0\ \, 0\ \, 0\ \, 0\ \, 1
\\
0\ \, 0\ \, 0\ \, 0\ \, 0\ \, 0\ \, 0\ \, 0\ \, 0\ \, 0\ \, 0\ \, 0\ \, 0\ \, 0\ \, 0\ \, 0\ \, 0\ \, 0\ \, 0\ \, 0\ \, 0\ \, 0\ \, 0\ \, 0\ \, 0\ \, 1\ \, 0\ \, 0\ \, 0\ \, 0\ \, 1\ \, 1
\\
0\ \, 0\ \, 0\ \, 0\ \, 0\ \, 0\ \, 0\ \, 0\ \, 0\ \, 0\ \, 0\ \, 0\ \, 0\ \, 0\ \, 0\ \, 0\ \, 0\ \, 0\ \, 0\ \, 0\ \, 0\ \, 0\ \, 0\ \, 0\ \, 0\ \, 0\ \, 1\ \, 0\ \, 0\ \, 1\ \, 1\ \, 0 
\\
0\ \, 0\ \, 0\ \, 0\ \, 0\ \, 0\ \, 0\ \, 0\ \, 0\ \, 0\ \, 0\ \, 0\ \, 0\ \, 0\ \, 0\ \, 0\ \, 0\ \, 0\ \, 0\ \, 0\ \, 0\ \, 0\ \, 0\ \, 0\ \, 0\ \, 0\ \, 0\ \, 1\ \, 1\ \, 1\ \, 0\ \, 0  
\end{array}
\right)
\label{bmatrixdbc}
\end{align}
Note that the only difference between (\ref{bmatrixdbc}) for the graph of  Fig. \ref{fig4} (b) and the corresponding matrix for  Fig. \ref{fig4} (a) is in the $B_{33}$ entry, denoted in boldface. For   Fig. \ref{fig4} (a) 
$B_{33}=1$, while $B_{33}=0$ for  Fig. \ref{fig4} (b) due to the defect. All other matrix elements are the same.

In the case of  Fig. \ref{fig4} (a), the characteristic polynomial $P(\lambda)$ for $B$ reads 
\begin{align}
P(\lambda)&=\lambda^{32}-8\lambda^{31}-24\lambda^{30}+338\lambda^{29}
 -79\lambda^{28}       -5838\lambda^{27}        +8344\lambda^{26}   +53206\lambda^{25} \nonumber \\
&\qquad   -118611 \lambda^{24}   -272276\lambda^{23}      +850672  \lambda^{22}    +731568 \lambda^{21}   
		-3608815 \lambda^{20}  \nonumber \\
&\qquad 		-523114 \lambda^{19}
     +9483060\lambda^{18}   
		  -2618090\lambda^{17}   -15501035\lambda^{16}     +8629236\lambda^{15} \nonumber    \\
&\qquad 		+15387876 \lambda^{14}   -11961756 \lambda^{13}    -8727651 \lambda^{12}
  +8860176\lambda^{11}     +2415600 \lambda^{10}  \nonumber \\
&\qquad   -3569022 \lambda^9    
-116235 \lambda^8      +727542 \lambda^7     -78732 \lambda^6     -58320 \lambda^5
      + 11664\lambda^4, 
\label{polyp}
\end{align}
with the characteristic polynomial of $A$ being ${\mathcal P}(\lambda)=P(\lambda)P(-\lambda)$. The roots of 
${\mathcal P}(\lambda)$ are given by (\ref{formulaobc}) with $L=8$. 

In the case of  Fig. \ref{fig4} (b), the characteristic polynomial $Q(\lambda)$ for $B$ reads 
\begin{align}
Q(\lambda) &=
           \lambda^{32}          -7 \lambda^{31}        -31 \lambda^{30}     
						+309 \lambda^{29}        +217 \lambda^{28}      -5679 \lambda^{27}       +3187 \lambda^{26}
       +56798\lambda^{25}   \nonumber \\ 
&\qquad 				-70399 \lambda^{24}   -338714 \lambda^{23}     +587333 \lambda^{22}  
				+1234858 \lambda^{21}   -2759107\lambda^{20}  \nonumber \\
&\qquad					-2672048 \lambda^{19} 
     +7975967\lambda^{18}     +2956747 \lambda^{17}  -14533255 \lambda^{16} \nonumber   \\ 
&\qquad		-299377  \lambda^{15}  +16620705\lambda^{14}   
		 -3246561\lambda^{13}   -11634583 \lambda^{12} \nonumber \\
&\qquad    + 3872804\lambda^{11}     +4737597 \lambda^{10}   -2021218 \lambda^{9} -1010163 \lambda^{8} 
\nonumber \\  
&\qquad			+512874 \lambda^7      +85077  \lambda^6    -55422 \lambda^5 
      +   324 \lambda^4      + 1296 \lambda^3  ,
\label{polyq}
\end{align}
and the characteristic polynomial of $A$ is $Q(\lambda)Q(-\lambda)$. Use of the computer algebra package MAGMA, hosted by The University of Sydney \cite{magma}, indicates that (\ref{polyq}) is not solvable in the Galois sense. That is, the roots cannot be expressed in terms of radicals. This is in stark contrast to the open $8\times 8$ lattice without defect. The roots of (\ref{polyp}) admit a closed form contained in the set  given by (\ref{formulaobc}) with $L=8$. Moreover, these roots are all expressible in terms of radicals, as a result of $x=\cos(\pi/9)$ being a solution of the cubic equation 
\begin{align*}
8x^3-6x-1=0.
\end{align*}
This indicates that the Bethe Ansatz solutions associated with  the lattices depicted in Fig. \ref{fig4} should not be considered equivalent in terms of their ``exactness''; i.e. being expressed in closed-form. To obtain the roots of (\ref{polyq}) requires numerical approximation, while obtaining the roots of (\ref{polyp}) does not. However, there is nothing apparent to suggest that they possess inequivalent types of integrability; both admit a set of commuting operators given by (\ref{con1},\ref{con2}), the formulae for which do not depend on knowledge of the eigenvalues of the submatrix $B$ of the adjacency matrix $A$. 

\section{Discussion}

In this work the Yang-Baxter integrable Hamiltonian (\ref{ham}), with conserved operators (\ref{con1},\ref{con2}),
was interpreted as a square lattice model with global-range interaction. The square lattice structure was obtained by a suitable choice of adjacency submatrix $B$, which was shown to be amenable to open, cylindrical, and toroidal boundary conditions. A Bethe Ansatz solution was presented that is dependent on the eigenvalues of $B$. From this solution expressions for ground-state energies in the case of $N=2$ (\ref{nrg2}) and $N=3$ (\ref{nrg3}) were obtained for arbitrary $L$, as were the associated imbalance fluctuations 
(\ref{i2},\ref{i3}). The $L$-dependence of these formulae is fully explicit through (\ref{high}).  

The matrix $B$ can be chosen to have more general forms. As an example, the case of a boundary defect was discussed in an open boundary $8\times 8$ lattice. This example serves to highlight what might be considered counterintuitive properties of the general model.  To obtain the energy eigenvalues requires both the diagonalisation of the matrix $B$, and the application of the Bethe Ansatz. The diagonalisation of the matrix $B$ is independent of any Bethe Ansatz calculation. For the $8\times 8$ lattice, the eigenvalues are given by the roots of a polynomial of order 32, $P(\lambda)$ given by (\ref{polyp}) for the open boundary case, and 
$Q(\lambda)$ given by (\ref{polyq}) for the model with a defect. In the former case, all roots are known in closed form; not so for the latter.
These eigenvalues of $B$ appear in the Bethe Ansatz equations. For a 3 particle system, the ground-state energy was identified whereby, via a change of variable, the roots of the Bethe Ansatz equations relate to the roots of polynomial of order 3, viz. $p(\lambda)$ given by (\ref{cubic}).  In obtaining the ground-state energy for a 3-particle system on an $8\times 8$ lattice with defect, the more demanding computational task is the diagonalisation of the matrix $B$, not obtaining the explicit solution of the Bethe Ansatz equations leaing to the energy (\ref{nrg3}).    

For arbitrary lattice sizes, the characteristic polynomial of $B$ has order $L^2/2$. Instinctually, at some value of $L$  the numerical evaluation of the eigenvalues of $B$ for a system with randomly chosen integrable defects will become intractable. On the other hand, the closed form (\ref{formulaobc}) for defect-free open boundaries is valid for all $L$. It is similarly the case for cylindrical (\ref{formulacbc}) and toroidal (\ref{formulatbc}) boundary conditions. These observations contend that within the class given by (\ref{ham}) there are some Hamiltonians for which an exact, 
closed-form expression for the Bethe Ansatz solution (\ref{nrg},\ref{bae}) is known. There are some Hamiltonians for which an exact, closed-form expression for the Bethe Ansatz solution (\ref{nrg},\ref{bae}) cannot be known.  These intuitive considerations are consistent with the formalist approach taken in \cite{l21} to conclude that Yang-Baxter integrability is not synonymous with exact-solvability. 

\section*{Appendix}
\def\aa{\mathfrak a}
\def\bb{\mathfrak b}
For the one-dimensional free-boson Hamiltonian on $L$ lattice sites with open boundary conditions 
\begin{align}
H=   \sum_{k=1}^{L-1} (\bb^\dagger_k \bb_{k+1} + \bb^\dagger_{k+1} \bb_k),
\label{h1}
\end{align}
the single-particle energies are known to be given by 
\begin{align}
E=2\cos\left( \frac{\pi p}{L+1} \right), \qquad \qquad p=1,\dots, L.
\label{open}
\end{align}
This result is obtained by substituting the canonical transformation
\begin{align*}
\bb_k&=\sqrt{\frac{2}{L+1}} \sum_{j=1}^{L} \sin\left(\frac{\pi j k}{L+1} \right) \aa_j, 
\\
\bb^\dagger_k&=\sqrt{\frac{2}{L+1}} \sum_{j=1}^{L} \sin\left(\frac{\pi  j k}{L+1} \right) \aa^\dagger_j
\end{align*}
into (\ref{h1}), bringing it into a diagonalised form. 

If periodic boundary conditions are imposed, the Hamiltonian reads 
\begin{align}
H=  \bb^\dagger_1 \bb_{L} + \bb^\dagger_{L} \bb_1 + \sum_{k=1}^{L-1} (\bb^\dagger_k \bb_{k+1} + \bb^\dagger_{k+1} \bb_k),
\label{h2}
\end{align}
and the single-particle energies are known to be given by 
\begin{align}
E=2\cos\left( \frac{2\pi p}{L} \right), \qquad p=1,\dots,L.
\label{closed}
\end{align}
This result is obtained by substituting the canonical transformation (where $i=\sqrt{-1}$)
\begin{align*}
\bb_k&=\sqrt{\frac{1}{L}} \sum_{j=1}^{L} \exp\left(\frac{2\pi i j k}{L} \right) \aa_j, 
\\
\bb^\dagger_k&=\sqrt{\frac{1}{L}} \sum_{j=1}^{L} \exp\left(\frac{-2\pi i  j k}{L} \right) \aa^\dagger_j
\end{align*}
into (\ref{h2}). 

Both of these examples can be expressed through the notion of an $L\times L$ adjacency matrix $A$, where $A_{jk}=1$ if there is hopping between the sites labelled $j$ and $k$, and $A_{jk}=0$ otherwise. Extension to an $L\times L$ square lattice, with an associated $L^2\times L^2$ adjacency matrix is accommodated via tensor products. Let $A_{\rm vert}$ denote the adjacency matrix for a one-dimensional lattice in the vertical direction, and let $A_{\rm horiz}$ denote the adjacency matrix for a one-dimensional lattice in the horizontal direction. The boundary conditions for $A_{\rm vert}$ and $A_{\rm horiz}$ may be chosen independently. The square lattice adjacency matrix is 
\begin{align*}
A=A_{\rm vert}\otimes {\mathbf I} + {\mathbf I} \otimes A_{\rm horiz}
\end{align*}
where ${\mathbf I}$ is the $L\times L$ identity matrix. Since  $A_{\rm vert}\otimes {\mathbf I}$ and 
${\mathbf I} \otimes A_{\rm horiz}$ commute, they are simultaneously diagonalisable. Thus the spectrum of $A$ is the sum of the parts provided by (\ref{open},\ref{closed}), leading to the expressions (\ref{formulaobc}), (\ref{formulacbc}) and (\ref{formulatbc}). 

\section*{Acknowledgments}
This research was supported by the Australian Research Council through Discovery Project DP200101339, {\it Quantum control designed from broken integrability}. I acknowledge the traditional owners 
of the Turrbal and Jagera country on which The
University of Queensland (St. Lucia campus) operates.
Part of this research was developed at the Mathematical Research Institute MATRIX, Creswick, Australia, in July 2024 during the research program {\it Mathematics and Physics of Integrability} (MPI2024). I thank Angela Foerster for her thoughtful, enthusiastic, and optimistic discussions during that meeting.

\end{document}